\documentclass[twocolumn,english,prb,notitlepage,superscriptaddress,nobibnotes]{revtex4-2}
\usepackage[T1]{fontenc}
\usepackage[latin9]{inputenc}
\usepackage{graphicx}
\usepackage{amsmath}
\usepackage{amssymb}
\usepackage{float}
\usepackage[colorlinks = true,
linkcolor = blue,
urlcolor  = blue,
citecolor = blue,
anchorcolor = blue,
pdftitle={YbCuxSe2_Random_Singlet}]{hyperref}
\usepackage[usenames, dvipsnames]{color}

\makeatletter
\usepackage{babel}
\usepackage{float}
\usepackage{appendix}

\usepackage{tabularx}
\newcolumntype{C}{>{\centering\arraybackslash}X}

\makeatother

\begin{document}

%
\title{Random singlet physics in exchange disordered 2D triangular YbCu$_{1.14}$Se$_2$}

\author{Caitlin S.T. Kengle}
\affiliation{MPA-Q, Los Alamos National Laboratory, Los Alamos, NM 87545, USA}

\author{S. M. Thomas}
\affiliation{MPA-Q, Los Alamos National Laboratory, Los Alamos, NM 87545, USA}

\author{R. Movshovich}
\affiliation{MPA-Q, Los Alamos National Laboratory, Los Alamos, NM 87545, USA}

\author{Shengzhi Zhang}
\affiliation{National High Magnetic Field Laboratory, Los Alamos National Laboratory, Los Alamos, NM 87545, USA}

\author{Eun Sang Choi}
\affiliation{National High Magnetic Field Laboratory, Tallahassee, Florida 32310, USA}

\author{Minseong Lee}
\affiliation{National High Magnetic Field Laboratory, Los Alamos National Laboratory, Los Alamos, NM 87545, USA}

\author{P.F.S. Rosa}
\affiliation{MPA-Q, Los Alamos National Laboratory, Los Alamos, NM 87545, USA}

\author{A. O. Scheie}
\affiliation{MPA-Q, Los Alamos National Laboratory, Los Alamos, NM 87545, USA}

\date{\today}

\begin{abstract}

		Quantum spin liquid (QSL) phases exist in theory, but real candidate QSL materials are often extraordinarily sensitive to structural defects which disrupt the ground state.  Here, we investigate candidate triangular QSL material YbCu$_{1.14}$Se$_2$ and discover the absence of magnetic order, but also no compelling evidence of a QSL ground state due to significant structural disorder. We instead look at the results through a lens of a 2-dimensional (2D) random singlet phase.  We are able to match thermodynamic measurements using a phenomenological model of a distribution of singlet formation. YbCu$_{1.14}$Se$_2$ behaves strikingly similar to other disordered triangular lattice materials, suggesting universal behavior of random singlet formation in 2D frustrated systems. 
		
\end{abstract}

\maketitle

\section{Introduction}

In 1973, Anderson predicted the ground state of the nearest-neighbor spin $=1/2$ Heisenberg triangular lattice antiferromagnet to be a quantum spin liquid (QSL) \cite{Anderson1973}: a unique phase of matter where magnetic frustration produces a long-range entangled state with exotic gauge fields and no magnetic order \cite{savary2016quantum,broholm2020quantum,Li_2020_triangular,RevModPhys.89.025003}. 
Anderson's prediction was later verified numerically: although the nearest neighbor does magnetically order, the triangular lattice Heisenberg antiferromagnet does indeed form a qsl with a weak second-neighbor exchange \cite{PhysRevB.93.144411,Gallegos2025,PhysRevB.92.041105,PhysRevB.92.140403,PhysRevB.96.075116}. 
Over the last few decades there has been an intense search for experimental realizations \cite{broholm2020quantum,Knolle_FieldGuide}. 
However, despite many candidate compounds, the QSL state in most materials are controversial because of defects and disorder disrupt the ground state \cite{Wen2019,Knolle_FieldGuide}. This obviously confounds the experimental search for QSLs, but it also raises an interesting question: what kind of state are ``failed QSLs'', where disorder in triangular lattices produces a fluctuating ground state without magnetic order? 

One of the most famous triangular QSL candidates is YbMgGaO$_4$ \cite{shen2016evidence,Paddison2017,Li2015,Xu_2016_YMGO,Zhang_2018_YMGO,rao2021survival_YMGO}. This compound has a 2D triangular lattice of magnetic Yb ions with intrinsic site disorder in an adjacent non-magnetic layer \cite{Li2015}, no magnetic ordering transition \cite{Paddison2017}, and sub-linear low-temperature specific heat \cite{Xu_2016_YMGO,Paddison2017}. Although originally interpreted to be a QSL \cite{shen2016evidence,Paddison2017,Li2015}, subsequent studies showed substantial exchange randomness \cite{Li_2017_YMGO}, frequency-dependent glass freezing around 100~mK \cite{Ma2018}, and no QSL quasiparticles in thermal conductivity \cite{Xu_2016_YMGO} (though this remains controversial \cite{rao2021survival_YMGO}). This led to a reinterpretation of YbMgGaO$_4$ as a magnetic disordered ground state \cite{Zhu_2017_YMGO,Kimchi_2018_YMGO}. 
And yet, YbMgGaO$_4$ is not simply a spin glass as it has substantial quantum fluctuations \cite{Paddison2017,shen2016evidence} and no residual entropy in its ground state \cite{Paddison2017}. What sort of phase then does YbMgGaO$_4$ form, and how universal is it?

A proposed answer to this question is a 2D version of a random singlet phase \cite{PhysRevB.45.2167,PhysRevB.50.3799}: where a disordered exchange landscape produces a frozen network of spin singlets (with corresponding singlet quantum fluctuations) \cite{Kimchi_2018_YMGO}. Similarly to the QSL phase, there is strong entanglement and no static magnetism \cite{shimokawa2025can}, but without the key feature of exotic collective quasiparticle excitations. The random singlet phase has been proposed as the ground state of a disordered spin liquid in antiferromagnetic Heisenberg triangular, Kagome, honeycomb, and square lattices \cite{PhysRevB.102.054443,Wu2019PRB, Kawamura2014JPSJ,Uematsu2017JPSJ,Shimokawa2015PRB,Uematsu2018JPSJ,Kimchi_2018_YMGO,PhysRevB.102.094407}. 

\begin{figure}
    \centering
    \includegraphics[width=0.9\linewidth]{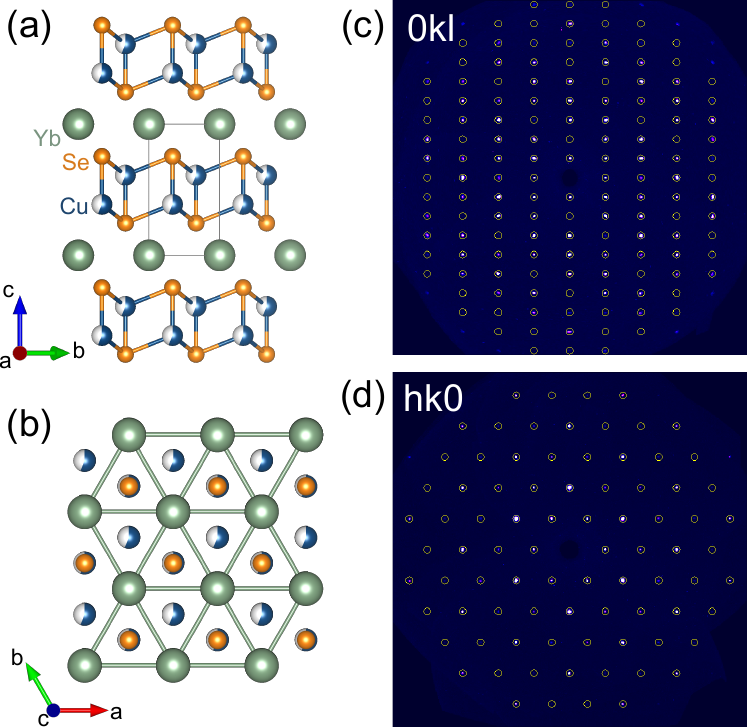}
    \caption{(a,b) Crystal structure of YbCu$_{1.14}$Se$_2$ obtained from SC-XRD measurements. The magnetic Yb ions sit at the corners of the unit cell, forming a triangular lattice. The copper position is shown as a partially filled sphere representing its partial occupancy. (c,d) $0kl$ and $hk0$ precession images. There are no deviations from the expected Bragg peak positions for a hexagonal $P$ lattice (yellow circles), indicating that the Cu vacancies are not ordered.}
    \label{fig:SCXRD_fig}
\end{figure}

Subsequent to  YbMgGaO$_4$, the $A$YbSe$_2$ [$A$=Na,K,Cs] series has emerged as promising QSL candidates \cite{bordelon2019field,Dai_2021,Scheie2024_KYS,Scheie2024_KYSNYS,xie2023complete}. 
Here we sought to investigate the related compound YbCu$_x$Se$_2$, wherein $x=1$ and Cu is substituted for the alkali ion in attempt to lower the insulating gap, potentially driving the material closer to a metal having a QSL ground state. 
This form of YbCu$_x$Se$_2$ has Yb$^{3+}$ ions in octahedral Se environments arranged in a perfect 2D triangular lattice pattern just like the $A$YbSe$_2$ series (and unrelated to zig-zag chain YbCuS$_2$ which belongs to space group $P2_12_12_1$ \cite{Ohmagari_2020,hori2023gapless}).
Our synthesis resulted in consistently off-stoichiometric crystals of YbCu$_x$Se$_2$, where $x=1.14$ which remain semiconducting despite the excess Cu (See Appendix \ref{sec:resist}).
We observe evidence of a spin disordered ground state with sublinear specific heat and a visible freezing transition at $T_f=0.1$~K in AC susceptibility. This behavior is strikingly similar to YbMgGaO$_4$ and matches theoretical quantum simulations of disordered triangular magnets, suggesting possible universal behavior of disordered QSL materials.

\section{Experiments and Results}

\subsection{Synthesis}
We synthesized single crystals of YbCu$_{1.14}$Se$_{2}$ using the self-flux technique~\cite{rosa2019flux}. We loaded Yb (Ames Laboratory, $> 99.9$\%), Cu (99.999\%), and Se (99.999\%) pieces in a 1:1:20 ratio into an alumina crucible and sealed everything under vacuum in a quartz ampule.
The reagents were slowly heated to 850~$^{\circ}$C, held at 850~$^{\circ}$C for 8~h, and slow cooled at 2~$^{\circ}$C/h to 600~$^{\circ}$C.
The ampule was then inverted, and the flux was removed $via$ centrifugation. The resultant plate-like crystals were up to 1~mm on a side and 0.2~mm thick.
Scanning electron microscope measurements reveal pristine surfaces mostly free of residual Se flux, and energy dispersive x-ray data yield a stoichiometry of YbCu$_{1.1(2)}$Se$_{1.8(3)}$.



\subsection{X-ray diffraction}
We performed single crystal x-ray diffraction (SC-XRD) on a $45\, \mu$m$\times35\,\mu$m$\times5\, \mu$m sample of YbCu$_{1.14}$Se$_2$. The unit cell was found to be $P$-$3m1$ ($\#$164) with lattice parameters $a = 4.0209(3)$~\AA$\:$ and $c = 6.4492(12)$~\AA$\:$, shown in Fig.~\ref{fig:SCXRD_fig}(a). 
The refinement reveals that the magnetic Yb ions form a triangular lattice which lies between polyhedral layers of nonmagnetic Cu and Se, shown in Fig.~\ref{fig:SCXRD_fig}(b). 

Copper occupies the 2d Wyckoff position, as shown in Fig.~\ref{fig:SCXRD_fig}(a), therefore the ``ideal'', i.e., stoichiometric, version of this compound should have non-ordered 50$\%$ occupation of the Cu site \cite{Daszkiewicz_2008}.
We find that the occupancy of Cu is 56.6$\pm$0.7\%, slightly above the expected value of 50\%, giving the compound a stoichiometry of YbCu$_{1.14}$Se$_2$. 
The $\sim57\%$ occupancy was confirmed in multiple samples, each with refinement residuals $<5\%$, and is consistent with other similar compounds in this series \cite{Esmaeili2014,Boswell2025ACS}.
The Yb and Se sites are fully occupied, and all occupations measured are consistent with the EDX measurements, within error.
All of the observed diffraction peaks correspond to the predicted Bragg positions of the unit cell, represented by yellow circles in Fig.~\ref{fig:SCXRD_fig}(c,d), indicating that the Cu vacancies are not ordered.
Atomic positions, isotropic displacement parameters, occupancies, and refinement details are presented in Table \ref{tab:atm_pos}.

\subsection{Bulk measurements}

\subsubsection{Magnetic susceptibility}

\begin{figure}
    \includegraphics[width=\columnwidth]{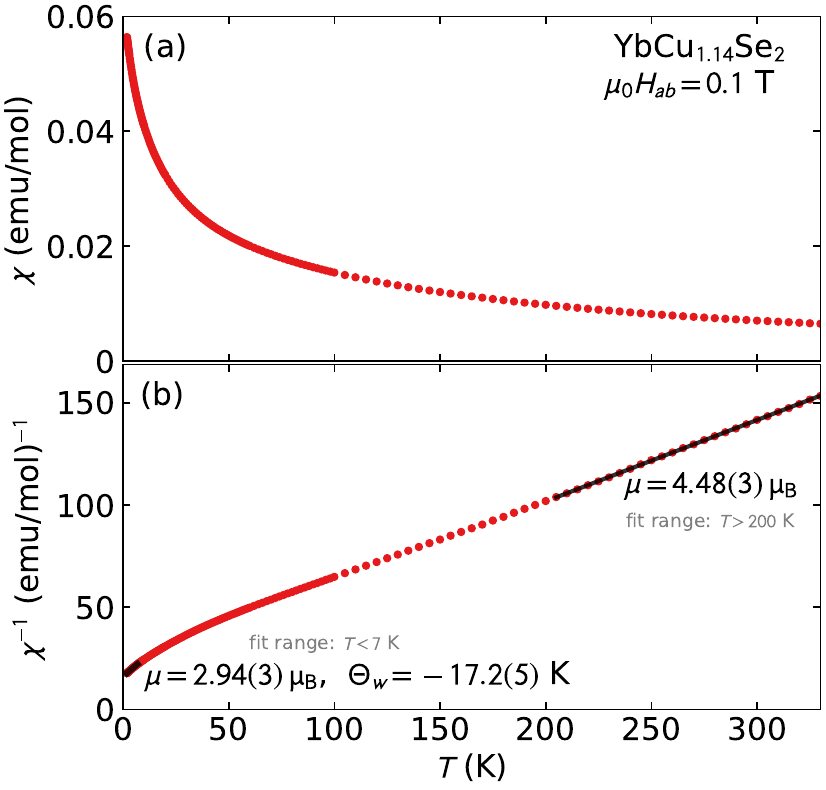} 
    \caption{YbCu$_{1.14}$Se$_2$ magnetic susceptibility as a function of temperature. Panel (a) shows the susceptibility, and panel (b) shows inverse susceptibility with Curie-Weiss fits to high-temperature and low-temperature data. The nonlinearity in $\chi^{-1}$ is due to crystal electric field effects.}
    \label{fig:susceptibility}
\end{figure}

We measured the temperature-dependent susceptibility of YbCu$_{1.14}$Se$_2$ in a Quantum Design MPMS between 2~K and 350~K in a 0.1~T in-plane magnetic field using a 0.2~mg crystal. Data are shown in Fig.~\ref{fig:susceptibility}. The susceptibility diverges at low temperatures consistent with a Kramer's doublet ground state. The inverse susceptibility [Fig.~\ref{fig:susceptibility}(b)] shows a nonlinearity that is almost certainly due to crystal electric field effects \cite{mugiraneza2022tutorial}. Comparison to KYbSe$_2$, which has a similar bend in susceptibility at slightly higher temperatures from a 17 meV crystal electric field (CEF) mode \cite{Xing2021KYS,Scheie2024_KYS}, suggests $\sim 10$~meV separating the first excited CEF level from the ground state doublet in YbCu$_{1.14}$Se$_2$.  

Fitting a Curie-Weiss law to the susceptibility $T>200$~K gives an effective moment of 4.48(3)~$\mu_B$, close to the free Yb$^{3+}$ moment of 4.54~$\mu_B$. Fitting the susceptibility $T<8$~K gives an effective moment of 2.94(3)~$\mu_B$ and a Weiss temperature $\Theta_W = -17.2(5)$~K, indicating antiferromagnetic exchange within the ground state doublet. Note, however, the magnitude of the low-temperature $\Theta_W$ should be taken cautiously, as the Weiss temperature comes from a mean field model that is only valid at high temperatures. Nevertheless, the overall sign of the exchange should be reliable. Meanwhile, the high temperature fit gives $\Theta_W = -54(4)$~K, but because CEF effects dominate this does not meaningfully indicate the magnitude or sign of magnetic exchange \cite{mugiraneza2022tutorial}.

\subsubsection{Specific heat}

\begin{figure*}
    \includegraphics[width=0.99\textwidth]{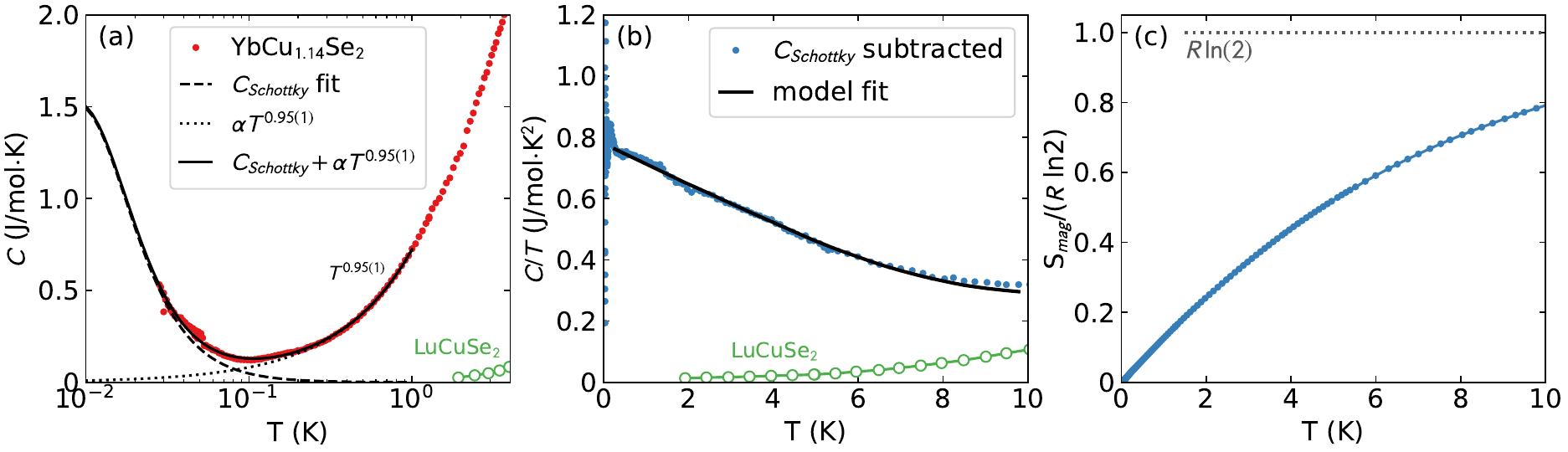} 
    \caption{Zero-field heat capacity of YbCu$_{1.14}$Se$_2$. Panel (a) shows the total measured temperature-dependent heat capacity ($C_{nuc} + C_{mag} + C_{phon}$) along with a modeled nuclear Schottky anomaly $C_{Schottky}$ and low-temperature electronic power law, and the specific heat from nonmagnetic analogue LuYbSe$_2$ \cite{khattar2025magnetic} as an estimate of the phonon contribution $C_{phon}$. Panel (b) shows $C/T$ with the nuclear Schottky upturn subtracted (effectively giving $C_{mag} + C_{phon}$) alongside the singlet-distribution model in Appendix \ref{sec:dimer-model}. The black line in panel (b) is the singlet-distribution model and the green data shows the LuCuSe$_2$ $C_{phon}$ phonon background (which was added to the model specific heat).  Panel (c) shows the integrated magnetic entropy (with the low-temperature nuclear Schottky upturn  and phonon contribution subtracted).}
    \label{fig:heatcapacity}
\end{figure*}

We performed specific heat measurements at zero field using a quasi-adiabatic thermal relaxation technique in two different cryostats.
First, we collected specific heat data from 10~K to 0.4~K in a Quantum Design Physical Property Measurement System (PPMS) cryostat equipped with a $^3$He insert. 
Second, we collected specific heat data between 2~K and 30~mK in an Oxford dilution refrigerator. 
The combined data are shown in Fig.~\ref{fig:heatcapacity}. 
The specific heat between 100~mK and 10~K shows no clear anomaly indicative of magnetic order, and besides a nuclear Schottky anomaly below 100~mK (which may not indicate static electronic magnetism because of Yb's substantial quadrupolar nuclear hyperfine interaction \cite{Bleaney_1963}), is largely featureless. In Fig.~\ref{fig:heatcapacity}(b) and (c) we model and subtract the nuclear Schottky anomaly \cite{scheie2019exotic} and plot $C/T$ and the integrated entropy $\Delta S = \int dT \frac{C}{T}$. 
The $C/T$ plot shows sub-linear specific heat from 100~mK up to 5~K, revealed primarily by the negative slope in Fig.~\ref{fig:heatcapacity}(b) for $T>0.2$~K (for $T<0.2$~K the behavior is ambiguous, see Appendix \ref{sec:dimer-model}).  Meanwhile, the integrated entropy between 100~mK and 10~K is around 80\%  $R\ln(2)$. Above 8~K the phonon contribution becomes significant, as indicated by the specific heat in nonmagnetic LuCuSe$_2$ \cite{khattar2025magnetic}, making it difficult to experimentally discern whether the magnetic entropy converges to $R\ln(2)$---though our theoretical models below indicate this is the case.

\subsubsection{Ultra-low-temperature susceptibility}

\begin{figure}
    \includegraphics[width=0.98\columnwidth]{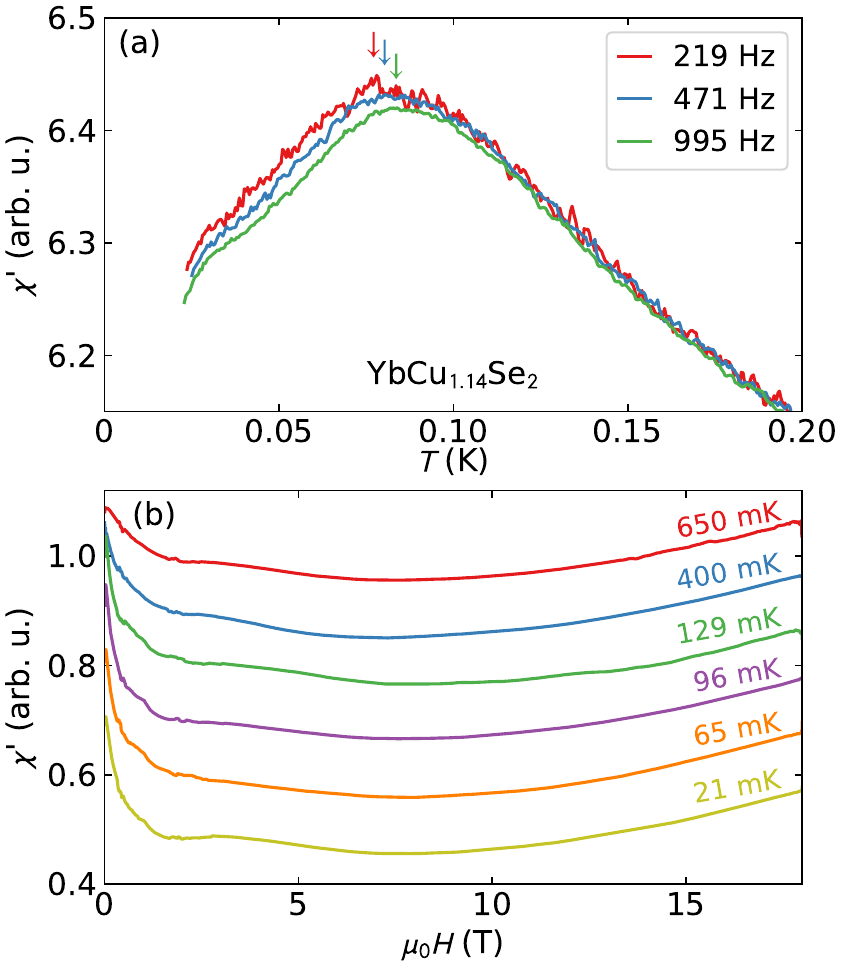} 
    \caption{Dilution refrigerator susceptibility of YbCu$_{1.14}$Se$_2$. Panel (a) shows the zero-field temperature-dependent susceptibility, showing a frequency-dependent symmetric peak indicating a glass-freezing transition around 80~mK. (b) shows the field-dependent susceptibility (offset vertically for clarity) at different temperatures. The absence of a sudden dip in susceptibility suggests YbCu$_{1.14}$Se$_2$ does not have a 1/3 magnetization plateau like $A$Yb$X_2$ compounds.}
    \label{fig:maglab}
\end{figure}

Finally, we measured the ultra-low-temperature magnetic susceptibility using a custom-built ac susceptometer at SCM1 of National High Magnetic Field Laboratory in Tallahassee, Florida. The susceptometer comprises a solenoidal coil that generates the ac excitation magnetic field and a pair of oppositely wound sensing coils positioned inside the solenoid. These sensing coils are designed to have equal but opposite mutual inductance, so that when connected in series, their induced voltages cancel, yielding a net zero signal in the absence of a sample. We placed a hexagon-shaped crystal, approximately 1 mm in its longest dimension, at the center of one of the sensing coils. The presence of the sample alters the magnetic flux, inducing a net voltage across the sensing coils that is proportional to the ac magnetic susceptibility. For more details, see ref.~\cite{lee2016magnetic}. Both the ac excitation field and the external dc magnetic field were applied along the in-plane direction. 
The data are shown in Fig.~\ref{fig:maglab}. 

In the zero-field temperature-dependent susceptibility, there is a symmetric frequency-dependent peak at $T_f \approx 0.1$~K which is characteristic spin-glass freezing behavior \cite{RevModPhys.58.801}. 
We find that the peak changes with applied field, shifting to higher temperatures and broadening, shown in Fig.~\ref{fig:chi_H} in Appendix \ref{sec:chi_app}.
In the constant temperature field-dependent susceptibility, the susceptibility curves are largely featureless.  Importantly, there is no dip in susceptibility from a 1/3 magnetization plateau observed in other Yb delafossites around 3-5~T \cite{bordelon2019field,Scheie2024_KYSNYS,belbase2025field}. This plateau is a key feature of the 2D Heisenberg triangular lattice model \cite{Nishimori_1986,Chubukov_1991}, and is present in most of the delafossite candidates for triangular QSL physics \cite{Xing2021KYS,bordelon2019field,Scheie2024_KYSNYS,xie2023complete}, but  disappears upon doping nonmagnetic sites into the lattice \cite{alvarado2025magneticdilutiontriangularlattice}. Its absence in YbCu$_{1.14}$Se$_2$ therefore signals a significant departure from the ideal 2D triangular Heisenberg model.

\section{Discussion}

	The SC-XRD results indicate a well-ordered triangular lattice of Yb ions with a significant amount of disorder on the Cu site: the Cu occupation is not the expected 50\%.
    The lack of ordering of Cu vacancies is consistent with previous x-ray diffraction measurements, where Cu occupancy was freely refined to be $1.08$ \cite{Boswell2025ACS}.
    However, the related compound DyCu$_x$Se$_2$ has a reported Cu occupancy of 57 \%, matching the our refined value remarkably well \cite{Esmaeili2014}. 
	Importantly, the disorder on the non-magnetic Cu itself is separated from the Yb site with Se, shown in Fig.~\ref{fig:SCXRD_fig}(a). 
    The disordered crystalline environment provides bond disorder which would create a distribution of magnetic exchanges, similar to that in YbZnGaO$_4$ and YbMgGaO$_4$ \cite{Ma2018, Kimchi_2018_YMGO}.
    
	Although the YbCu$_{1.14}$Se$_2$ susceptibility indicates a well-defined local Yb$^{3+}$ spin, heat capacity shows no clear magnetic ordering transition (no spike or discontinuity that is unambiguous). 
	Meanwhile, the sub-linear specific heat between 0.5~K and 5~K shows a manifestly non-magnon behavior.
	A sub-linear specific heat is neither consistent with a 2D bosonic magnon dispersion (which in the low-$T$ limit predicts $C \propto T^2$ for gapless linear modes or $C \propto e^T$ if gapped) nor with a Fermi surface (which regardless of dimensionality predicts $C \propto T$ as $T \rightarrow 0$) \cite{AshcroftMermin}. 
	Sub-linear specific heat 
    has been observed in exchange-disordered magnets like  Li$_4$CuTeO$_6$ \cite{Khatua2022} and YbMgGaO$_4$ \cite{Xu_2016_YMGO}; furthermore, theoretical calculations of a strongly disordered 2D triangular Heisenberg antiferromagnet predicts a slightly sub-linear specific heat \cite{Wantanabe_2014,Kimchi_2018_YMGO},
    and a bond-disordered Kitaev magnet also theoretically has sub-linear low temperature specific heat \cite{Knolle_2019_BondDisordered}. This suggests perhaps a universal feature of disordered frustrated systems.  

\begin{figure}[t]
    \centering
    \includegraphics[width=\linewidth]{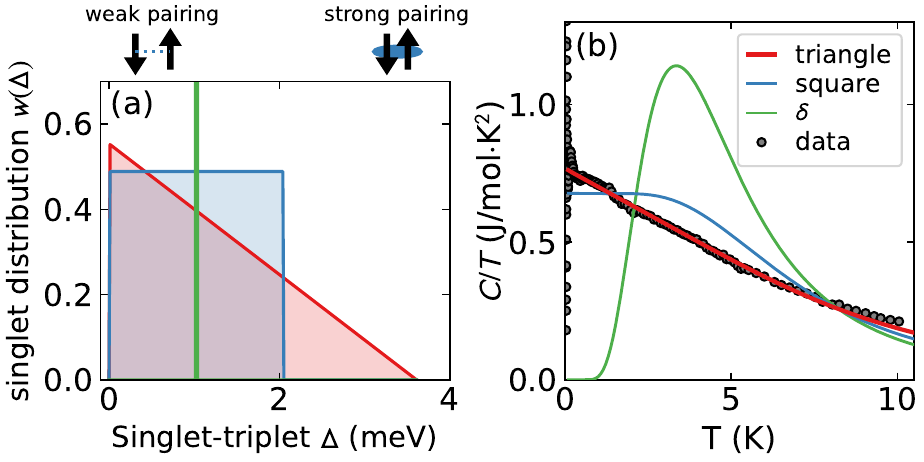}
    \caption{Calculated specific heat for distributions of singlets. Panel (a) shows the relative distribution $w(\Delta)$ of singlet-triplet gaps $\Delta$ for a triangular (red), square (blue) and  Dirac-delta function (green) distribution. Panel (b) shows the numerically computed heat capacity via Eq. \ref{eq:SumSpecificHeat}, shown overtop the nuclear  Schottky and phonon-subtracted experimental data from Fig.~\ref{fig:heatcapacity}. }
    \label{fig:DimerModel}
\end{figure}

    The specific heat in Fig.~\ref{fig:heatcapacity} shows a power law $C \propto T^{0.95(1)}$  up to 1~K, but above $T=1$~K the behavior is sublinear but not power law (Fig.~\ref{fig:PowerLawHC} in Appendix \ref{sec:dimer-model}). 
    To describe this behavior, we build a phenomenological specific heat model by summing over a distribution of $S=1/2$ singlet heat capacities of a system of disordered quantum singlets. We begin with the equation for specific heat of a generic multi-level system
\begin{equation}
    C(T) = \frac{1}{Z k_B T^2} \Big[
    \sum_i E_i^2 e^{\frac{-E_i}{k_B T}} -
    \frac{1}{Z} \big( \sum_i E_i e^{\frac{-E_i}{k_B T}} \big)^2
    \Big] \label{eq:GenericSpecificHeat}
\end{equation}
where $E_i$ are the energy levels (for singlet-triplet gap $\Delta$, $E = \{0, \Delta, \Delta, \Delta \}$) and $Z$ is the partition function \cite{AshcroftMermin}. 
We then compute the specific heat for a distribution $w(\Delta)$ of different $\Delta$ values weighted corresponding to the relative abundance of singlets with that splitting: 
\begin{equation}
C(T)_{sum} = \int  w(\Delta) C_{\Delta_i}(T)  \> d \Delta   \label{eq:SumSpecificHeat}
\end{equation} 
where Eq.~\ref{eq:GenericSpecificHeat} is divided by 2 to get the specific heat per spin rather than per dimer. 
Examples of three different $w(\Delta)$ distributions are shown in Fig. \ref{fig:DimerModel}: a square distribution, a triangular distribution, and a delta function (all singlet-triplet gaps identical).  Note that $\int w(\Delta) \> d \Delta = 1$ (for further details, see Appendix \ref{sec:dimer-model}).

	One peculiarity about the Eqs.~\eqref{eq:GenericSpecificHeat} and \eqref{eq:SumSpecificHeat} model is that the $T \rightarrow 0$ $C/T$ behavior matches the functional form of $w(\Delta)$ as $\Delta \rightarrow 0$. 
	In Appendix \ref{sec:dimer-model} we show that that this relationship is rigorous, and that 
	\begin{equation}
	    \lim_{T \rightarrow 0} \frac{C_{sum}}{T} \propto w(u_0 k_B T)
	\end{equation}
	where $w(u_0 k_B T)$ is the singlet distribution function with $\Delta$ replaced by a factor of $T$ (and $u_0 = 3.55978...$). 
	This means that the low-$T$ $C/T$ can be read directly as the low-$\Delta$ distribution of singlets.

	 Following the triangular low-$T$ $C/T$ in Fig.~\ref{fig:DimerModel}(b), we find that a triangular distribution of singlet gaps matches the data very well, which is plotted in Fig.~\ref{fig:DimerModel} (and also in  Fig.~\ref{fig:heatcapacity}(b) with  a non-magnetic LuCuSe$_2$ specific heat subtracted as a phonon contribution). The mean fitted singlet-triplet gap is $\langle \Delta \rangle = 1.20 \> {\rm meV} = 13.9$~K, which corroborates the energy scale from the low-temperature Weiss fit.
	This suggests a triangular distribution of singlet-triplet gaps:
	the density of spin pairs with linearly decreases with the magnitude of the pairing. In this linear distribution, 2.8\% of singlets have $\Delta > 3$~meV splitting, and nearly half (47\%) of singlets have $\Delta < 1$~meV splitting. 
	This is essentially the picture of random singlet formation, where the strong coupling in a random exchange model produces a small number of strong singlets, but as singlet pairs grow increasingly separated, the splitting grows increasingly numerous and weak. This simple toy model neglects many-body effects beyond pairwise singlet formation, but shows that the sublinear specific heat is consistent with a distribution of singlet pairing from Cu vacancy disorder. 
	
	AC susceptibility measurements indicate a spin freezing temperature of $\sim0.1$ K, similar to disordered triangular YbMgGaO$_4$ and YbZnGaO$_4$ \cite{Ma2018}.
	This frequency-dependent peak has been observed in other triangular Yb compounds \cite{Bag-Haravifard_2024,belbase2025field,scheie2024spectrum}, and can be generically interpreted as an energy scale below which the spins are frozen, which, due to the distribution of exchange energies, depends on the frequency probed. 
	The freezing energy scale here ($T_f \sim 0.1$~K) is very small compared with a 17.2(5)~K Weiss temperature (or the mean singlet-triplet splitting from heat capacity  $\langle \Delta \rangle/k_B = 13.9$~K), which suggests strong quantum dynamics above the freezing temperature. 
	
	Occam's razor would suggest that disorder destroys the possible QSL state in YbCu$_{1.14}$Se$_2$. However, the facts that (i) $T_f$ is two orders of magnitude below the $\sim 10$~K exchange energy scale from the Weiss temperature and specific heat fitted $\langle \Delta \rangle$, and (ii) that the majority of $R\ln(2)$ magnetic entropy is recovered above $T_f$, together indicate a significant amount of magnetic correlation buildup that is not glass-like. 
	Thus it could be that above a certain energy scale, there are substantial quantum dynamics and correlations present.
	Indeed, recent exact diagonalization simulations of a strongly disordered 2D triangular lattice reveal a nontrivial amount of quantum entanglement depth in the ground state \cite{shimokawa2025can}. I.E., the entanglement is not simply distributed pairwise, but forms a nontrivial longer-ranged structure. 
	This behavior is reminiscent of 2D square lattice random singlet phases \cite{PhysRevX.8.041040,PhysRevB.102.054443,PhysRevB.111.014409,PhysRevB.104.054201,PhysRevLett.127.017201}, but here we add the additional ingredient of a frustrated lattice geometry, which can enhance the entanglement in the system. 
	    
	The behavior of YbCu$_{1.14}$Se$_2$ is strikingly similar to YbMgGaO$_4$: intrinsic disorder in a non-magnetic layer adjacent to magnetic Yb ions \cite{Li2015}, no magnetic ordering transition \cite{Paddison2017}, sub-linear specific heat \cite{Xu_2016_YMGO,Paddison2017}, and frequency-dependent glass freezing around 100~mK \cite{Ma2018}. 
	This suggests a possible universal behavior of disordered triangular lattice magnets. 
	
	This universality may extend beyond Yb triangular lattices. Empirically, many frustrated magnets show extreme sensitivity to disorder: compare triangular $\kappa$-(BEDT-TTF)$_2$Cu$_2$(CN)$_3$ \cite{miksch2021gapped}, triangular Cs$_2$CuCl$_3$ \cite{PhysRevB.82.014421}, Kagome Herbertsmithite \cite{PhysRevLett.127.267202}, pyrochlore Yb$_2$Ti$_2$O$_7$\cite{PhysRevB.95.094407}, and frustrated square Sr$_2$CuTe$_{1-x}$W$_x$O$_6$  \cite{PhysRevLett.126.037201}. 
	Rather than characterize them as independent disordered or glassy states, they may be in a class of 2D random singlet-forming states. The addition of YbCu$_{1.14}$Se$_2$ to this family reinforces the generality of these features, and calls for a generic explanation of exchange-disordered quantum spin liquids. 

\section{Summary and Conclusion}
	We synthesized the 2D triangular YbCu$_{1.14}$Se$_2$ and showed that it had ideal 2D planes of magnetic Yb with intrinsic Cu disorder between planes. Magnetic susceptibility and heat capacity show no magnetic order and evidence of spin glass freezing, putatively driven by the distribution of magnetic exchanges from the disordered crystal structure. 
	
	The large difference between the Weiss temperature and spin freezing temperature $T_f$, as well as the significant magnetic entropy recovered above $T_f$, indicate substantial correlations above the freezing temperature. The sub-linear specific heat is also reminiscent of the disordered triangular Yb magnet YbMgGaO$_4$, which shows some very non-trivial quantum behaviors. Thus, despite the glassy features, YbCu$_{1.14}$Se$_2$ appears to show strong quantum behavior and is a candidate for studying 2D random singlet physics in a frustrated lattice.

	The abundance of failed QSL candidate materials opens the question: are disordered triangular spin liquids 2D random singlet phases? Magnetic frustration is a crucial component for enhancing both the QSL and random singlet phases. But where disorder destroys the QSL, it provides the environment for singlet pairing. Here, we show that 2D materials which have failed to show QSL behavior due to lattice disorder still show strong quantum behavior, and may have a universality class of their own.

\section*{acknowledgments}

This work was primarily supported by the Quantum Science Center (QSC), a National Quantum Information Science Research Center of the U.S. Department of Energy. 
Single crystal diffraction work by C.S.K. is supported by the U.S. Department of Energy, Office of Basic Energy Sciences, Division of Materials Science and Engineering under project "Quantum Fluctuations in Narrow-Band Systems."
M.L. acknowledges the support of the Laboratory Directed Research and Development program of Los Alamos National Laboratory. A portion of this work was performed at the National High Magnetic Field Laboratory, which is supported by the National Science Foundation Cooperative Agreement No. DMR-2128556, the State of Florida, and the Department of Energy.  Scanning electron microscope and energy dispersive x-ray measurements were performed at the Center for Integrated Nanotechnologies, an Office of Science User Facility operated for the U.S. Department of Energy, Office of Science. 

\section*{Data Availability}

The data and support the findings of this article are openly available \cite{data_availability}.

\appendix

\section{SC-XRD refinement details} \label{sec:SCXRD_app}

We collected the data for crystal structure refinement using a Bruker D8 VENTURE KAPPA single crystal x-ray diffractometer with an I$\mu$S 3.0 microfocus source (Mo K$\alpha$ $\lambda$ = 0.71073 \AA), a HELIOS optics monochromator, and a PHOTON II CPAD detector. We integrated the data with SAINT V8.41 and a multi-scan absorption correction was applied. Initial structural models were obtained via intrinsic phasing methods in SHELXT 2018/2 and refined by full-matrix least-squares methods against F$^2$ using SHELXL-2019/2 \cite{sheldrick2015crystal}.

\begin{table}[h]
\caption{Atomic positions, isotropic displacement parameters, and occupancies of YbCu$_{1.14}$Se$_2$.}
\centering
\begin{tabular}{c|c|c|c|c|c}
    \hline \hline
    Atom     & x        & y        & z         & Occ.      & U$_{iso}$ \\ \hline
    Yb       & 0        &  0       & 0         & 1         & 0.0129(2) \\ 
    Se       & 1/3      & 2/3      & 0.75406(12) & 1         & 0.0216(7) \\ 
    Cu       & 1/3      & 2/3      & 0.3802(3) & 0.566(7) & 0.0108(2)  \\ \hline \hline
\end{tabular}
\label{tab:atm_pos}
\end{table}

\begin{table}[h]
\caption{Single crystal refinement data for YbCu$_{1.14}$Se$_2$.}

\centering
\begin{tabular}{c|c}
    \hline \hline
    Temperature & 295 K \\ 
    Formula weight (g/mol) & 403.40 \\
    Space group, Z & $P$-$3m1$, 1 \\  
    a (\AA) & 4.0209(3) \\ 
    c (\AA) & 6.4492(12) \\ 
    V (\AA$^3$) & 90.30(2)\\
    Extinction coefficient & 0.138(7) \\ 
    Wavelength (\AA) & 0.71073 \\ 
    $\theta$ range ($^\circ$) &  3.159-30.418 \\ 
    No. reflections, R$_{int}$ & 3541, 0.0553 \\ 
    No. independent reflections &  131 \\ 
    No. parameters &  11 \\ 
    $R_1$, $wR_2$       & 0.0135, 0.332       \\ 
    GooF & 1.095 \\ \hline \hline
\end{tabular}
\label{tab:sc_parameters}
\end{table}

\section{AC susceptibility in field} \label{sec:chi_app}

\begin{figure}[h!]
    \centering
    \includegraphics[width=1\linewidth]{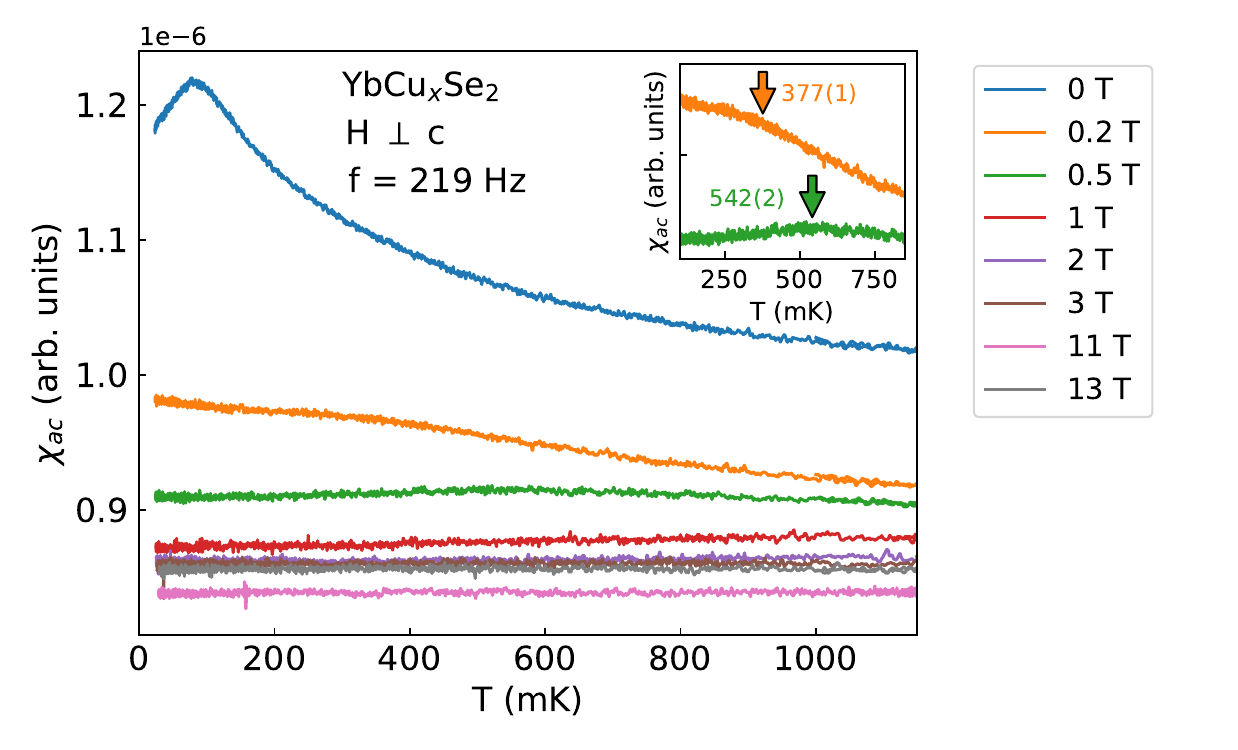}
    \caption{Susceptibility measurements at low temperature with applied field. The ac frequency was fixed to $f=219$ Hz and the magnetic field was applied in the plane. The inset shows a zoomed-in view of the $H=0.2$ and $0.5$ T measurements with the transition temperature marked.}
    \label{fig:chi_H}
\end{figure}

We performed dilution refrigerator AC susceptibility measurements as a function of temperature in the same manner described above using an ac frequency of 219 Hz and field applied in the plane.  The measurement protocol was to first sweep the field up to 18 T at a particular temperature, increase the temperature, then sweep down the field down to 0 T. Data are shown in Fig.~\ref{fig:chi_H}. The $H=0$ plot is the same as in Fig.~\ref{fig:susceptibility}(a) and shows the $T_f \simeq 0.1$ K spin freezing. With increasing field, the bump moves up in temperature and broadens, shown in the inset of \ref{fig:chi_H}. At $H=0.2$ T, the susceptibility has a significant linear portion plus a broad peak whereas by $H=0.5$ T, the curve is mostly flat with just a bump at $T_f$. At $H=0.2$ T, the bump is centered at 377(1) mK and at $H=0.5$ T it has reached 542(2) mK, obtained by fitting the broad peaks with Gaussian functions. The linear portion of the $H=0.2$ T data was subtracted off before fitting with the Gaussian function to obtain the center value.    

\section{Resistivity in zero field} \label{sec:resist}
\begin{figure}[h!]
    \centering
    \includegraphics[width=1\linewidth]{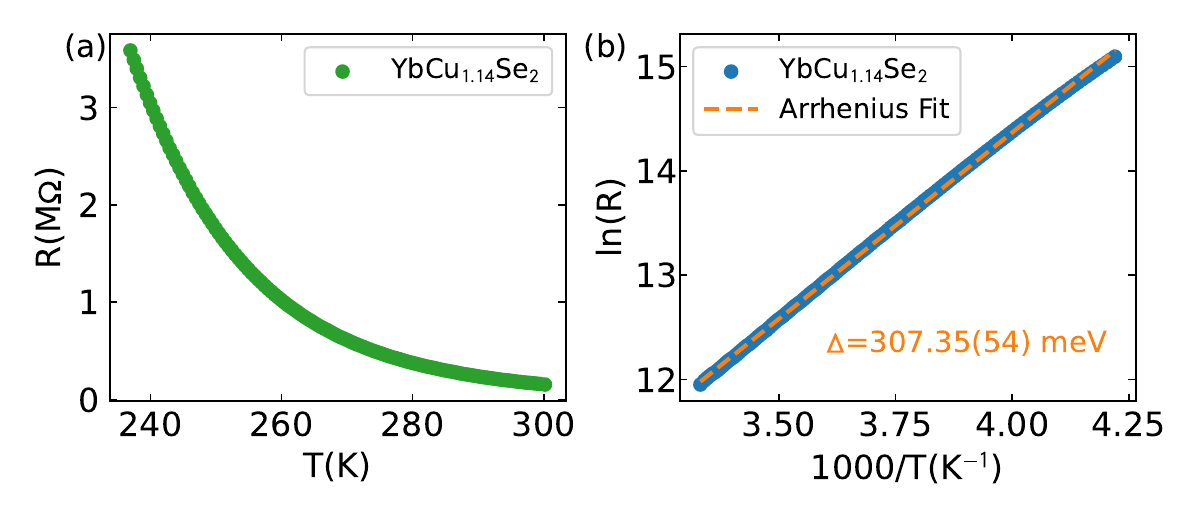}
    \caption{(a) Resistance versus temperature of YbCu$_{1.14}$Se$_2$ and (b) the resulting Arrhenius plot fit showing semiconducting behavior.}
    \label{fig:YbCuSe2_R_Arrhenius}
\end{figure}

Since Cu was chosen to push this material closer to a metallic state---and since there is an excess of Cu in the samples measured here---it is important to verify that the material remains insulating.
We performed resistance measurements on YbCu$_{1.14}$Se$_2$ and observe semiconducting behavior from 236 to 300 K [Fig.~\ref{fig:YbCuSe2_R_Arrhenius}(a)]. The Arrhenius fit taken from this data gives a gap value of 307.35(54) meV [Fig.~\ref{fig:YbCuSe2_R_Arrhenius}(b)]. Such behavior is consistent with other compositions of YbCu$_x$Se$_2$, which also show semiconducting behavior with resistivity \cite{Boswell2025ACS}.

\section{Nuclear Schottky model}
We modeled the nuclear Schottky anomaly using Eq.~\eqref{eq:GenericSpecificHeat} and energy levels defined by a linear Zeeman splitting 
\begin{equation}
\mathcal{H}_{zeeman} = a \langle J_z \rangle I_z
\end{equation}
where $J_z$ and $I_z$ are the electronic and nuclear spin operators and $a$ is a coupling constant taken from Ref.~\cite{Bleaney1963}. We computed the specific heat from the of the $I = 5/2$ and $I=7/2$ of $^{171}$Yb and $^{173}$Yb nuclei, with their natural abundance ratios of 14.2\% and 16.1\% respectively \cite{NIST_isotopes} to get the nuclear Schottky anomaly in the main text. We assumed a uniform static electronic moment $\langle J_z\rangle$ which fitted to a value of $0.779(9) \> {\rm \mu_B}$, which in principle corresponds to the root-mean-squared (RMS) static electronic moment \cite{scheie2019exotic}---however the substantial nuclear quadrupole coupling of Yb will also produce an energy splitting which produces a nuclear Schottky anomaly even with zero $\langle J_z\rangle$ \cite{Bleaney1963}, so this is not necessarily a reliable measure of the RMS static electronic moment. 
Nevertheless, this procedure matches the nuclear Schottky anomaly well and allows us to isolate the electronic heat capacity at low temperatures.

\section{Singlet-distribution specific heat model} \label{sec:dimer-model}

\begin{figure}
    \centering
    \includegraphics[width=0.85\linewidth]{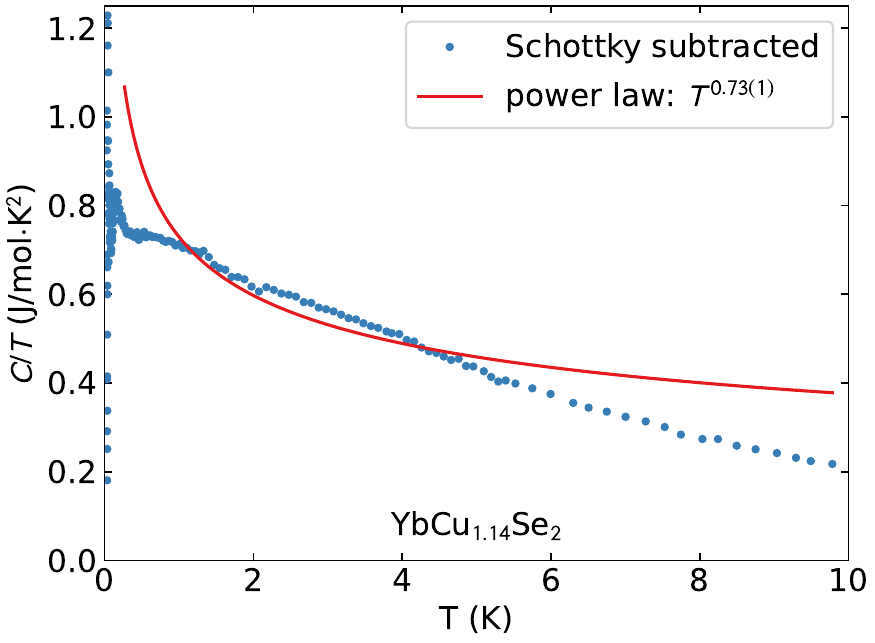}
    \caption{YbCu$_{1.14}$Se$_2$ zero-field magnetic specific heat (nuclear Schottky $C_{nuc}$ and phonon $C_{phon}$ contributions subtracted) fitted to a power law for 0.5~K$<T<$5~K (red line) and fitted to $T<$1~K (black dashed line, from Fig. \ref{fig:heatcapacity}[a]). As is evident from the fit, the trend of the data is linear in $C/T$ vs $T$, which does not follow a power law.}
    \label{fig:PowerLawHC}
\end{figure}

\begin{figure*}
    \centering
    \includegraphics[width=\linewidth]{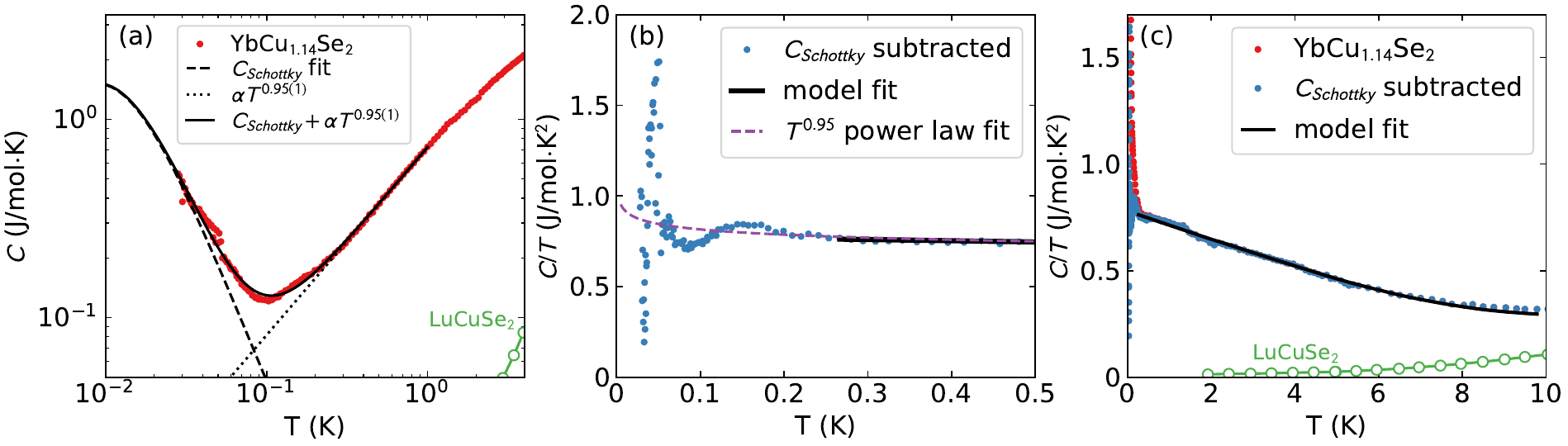}
    \caption{YbCu$_{1.14}$Se$_2$ zero-field total specific heat from Fig.~\ref{fig:heatcapacity} ($C_{nuc} + C_{mag} + C_{phon}$), plotted in a log-log plot  and the fitted power law (which fails above $T=1$~K, see Fig. \ref{fig:PowerLawHC}) (a) and with a close-up of $C/T<0.5$~K (b) ($C_{mag} + C_{phon}$). The plot in (b) shows significant noise in the data where the nuclear Schottky has been subtracted. There is also a ``hump'' in the data at 0.15~K potentially indicating correlations beyond the random singlet model. Panel (c) shows the full temperature range of specific heat with and without the nuclear Schottky contribution $C_{nuc}$, showing that the sublinear specific heat above $T=0.5$~K does not depend on how the nuclear Schottky anomaly is treated. }
    \label{fig:HC-closeup}
\end{figure*}

Despite the fact that the $T<1$~K data roughly follows a $T^{0.95(1)}$ power law, above 1~K the behavior strongly deviates from power law behavior. 
Figure \ref{fig:PowerLawHC} shows the zero field CuYbSe$_2$ specific heat fitted to a power law, which provides a poor match.  Fig.~\ref{fig:HC-closeup} shows the specific heat data focusing on the low-temperature behavior. This highlights the inexact fit of the $T^n$ power law below $T=0.2$~K. The small hump between 0.1~K and 0.2~K may be a real feature of the density of states, as the system approaches the glass freezing transition temperature from susceptibility and correlations extend beyond pairwise singlets. However, above $T=0.2$~K (where the nuclear Schottky anomaly is essentially negligible) the specific heat becomes featureless and monotonically decreasing with a negative slope in $C/T$. 
To better describe the specific heat over its full temperature range, we build a phenomenological model of disordered quantum singlets as described in the main text. 

\begin{figure}
    \centering
    \includegraphics[width=\linewidth]{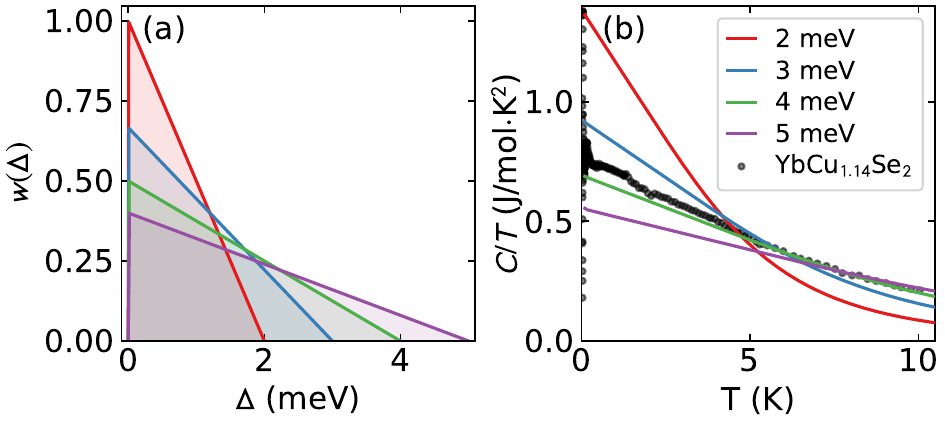}
    \caption{Calculated specific heat for triangular distributions of singlets, showing how the range of singlet distribution determines the $T \rightarrow 0$ slope of the computed specific heat. }
    \label{fig:DimerModel2}
\end{figure}

Remarkably, the specific heat computed from the triangular $w(\Delta)$ distribution matches the linear $C/T$ vs $T$ data extremely well between 1~K and 5~K. 
There is no physical reason why the distribution must be triangular, but we do note that this vaguely matches the expectation for random singlets: where a small number of singlets have strong pairing (large $\Delta$) and a large number of singlets have weak pairing (small $\Delta$). See the schematic in Fig.\ref{fig:DimerSusceptibility}(a). 
The negative slope of the $C/T$ vs $T$ curve is determined by the singlet distribution range in energy, shown in Fig.~\ref{fig:DimerModel2}. 
Fitting the range of the triangular distribution yields a best fit maximum gap of 3.611(7)~meV (the one free parameter in this model). When plotted atop a phenomenological $T^3$ Debye phonon contribution \cite{AshcroftMermin}, the model matches the data up to 10~K, as shown in Fig.~\ref{fig:heatcapacity}. 
The calculated specific heat has no residual entropy as it by construction converges to $R\ln(2)$. This indicates that there is in fact no residual magnetic entropy in YbCu$_{1.14}$Se$_2$, despite the disordered exchange and glass freezing.

\subsection*{$T \rightarrow 0$ behavior of $C/T$}

\begin{figure}
    \centering
    \includegraphics[width=0.85\linewidth]{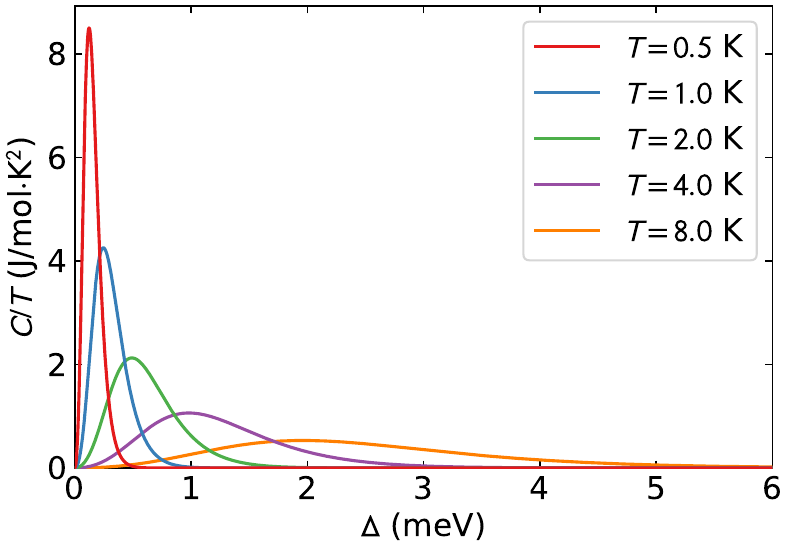}
    \caption{Calculated specific heat from Eq.~\eqref{eq:DimerHC} as a function of $\Delta$ for various temperatures $T$. As $T\rightarrow 0$, the function becomes sharper and narrower, and can be approximated by a Dirac delta function Eq.~\eqref{eq:deltafunc}.}
    \label{fig:DeltaFunction}
\end{figure}

Figures \ref{fig:DimerModel} and \ref{fig:DimerModel2} show a correspondence between $w(\Delta)$ and the low-$T$ functional form of $C/T$. 
We can mathematically show this relationship to be generic. 
Considering Eq.~\ref{eq:GenericSpecificHeat} for a dimer singlet with triplet gap $\Delta$, we simplify the equation to be 
\begin{equation}
    C(\Delta,T) = \frac{3 e^{\frac{\Delta}{k_B T}}\Delta^2}
    {2(3+e^{\frac{\Delta}{k_B T}})^2 k_B T^2}. 
    \label{eq:DimerHC}
\end{equation}
The functional form of this equation is shown in Fig.~\ref{fig:DeltaFunction}. In the limit that $T \rightarrow 0$, Eq.~\eqref{eq:DimerHC} becomes sharper and narrower and can be approximated as a Dirac delta function in $\Delta$. The integral of $C(\Delta, T)/T$ over $\Delta$ is a constant $A = \int_0^{\infty} d\Delta C(\Delta,T)/T =  1.93938 \> R \> k_B$ where $R$ is the gas constant and $k_B$ is the Boltzmann constant. Meanwhile, the peak center in $C(\Delta,T)/T$ in $T$ is found at $T = \frac{\Delta}{u_0 k_B}$ where $u_0 = 3.559780297...$ (a transcendental numerical root).  
Thus, in the $T \rightarrow 0$ limit we can write 
\begin{equation}
    \lim_{T \rightarrow 0} C(\Delta,T)/T = A \> \delta\left(u_0 k_B T - \Delta \right) 
    \label{eq:deltafunc}
\end{equation}
Inserting into Eq.~\eqref{eq:SumSpecificHeat} and carrying out the integral then gives 
\begin{equation}
    \lim_{T \rightarrow 0} \frac{C(T)_{sum}}{T} = A  \> w(u_0 k_B T). 
    \label{eq:approxCT}
\end{equation}
where the constants $A$ and $u_0$ are defined above.

\begin{figure*}
    \centering
    \includegraphics[width=\linewidth]{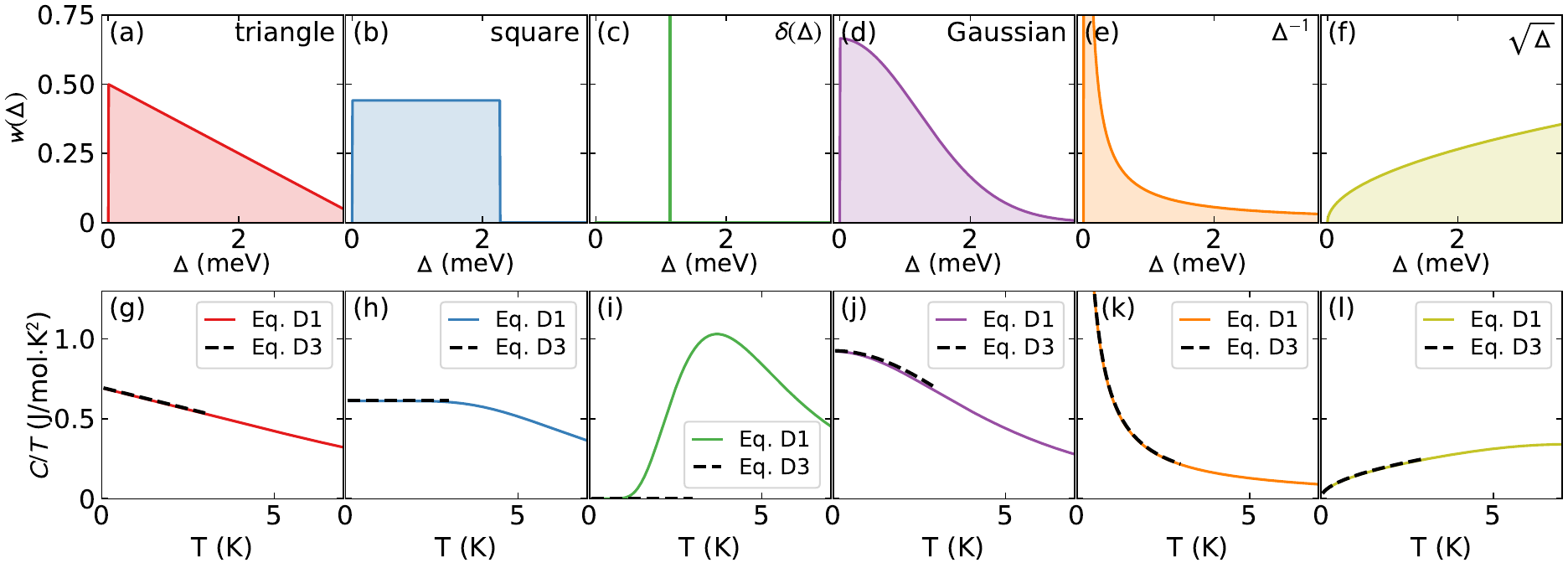}
    \caption{Calculated specific heat for six distributions of singlets(a)-(f), comparing the computed heat (g)-(l) from both Eqs.~\eqref{eq:DimerHC} and \eqref{eq:approxCT}. If $T$ is small compared to the energy scale of the distribution, the correspondence is near-perfect. Interestingly, for the power-law distributions in panels (e) and (f), the correspondence seems to hold up to higher temperatures as well.}
    \label{fig:DimerModel_differentDistributions}
\end{figure*}

This equation explicitly shows how in the $T \rightarrow 0$ limit, $C/T \propto w(u_0 k_b T)$ where $w(\Delta)$ is the distribution of singlets.    
This explains why the functional form of $w(\Delta)$ matches the low-$T$ $C/T$. 

Figure \ref{fig:DimerModel_differentDistributions} compares Eq.~\eqref{eq:approxCT} to Eq.~\eqref{eq:DimerHC} for a variety of $w(\Delta)$ distributions and shows that up to an appreciable $T$, Eq.~\eqref{eq:approxCT} is accurate.  (Note that we have only carried out this analysis for smoothly varying $w(\Delta)$ near $\Delta \rightarrow 0$; if this is not the case, the assumptions behind treating Eq.~\ref{eq:DimerHC} as a Dirac delta function may break down.) 
This implies that one can directly read the distribution of singlets from the low-$T$ behavior of $C/T$. For YbCu$_{1.14}$Se$_2$, the singlet distribution is clearly triangular (or something close to triangular) with a negative slope in $w(\Delta)$. 

Interestingly, for the power law distributions $w(\Delta) = \Delta^{-1}$ and $w(\Delta) = \Delta^{1/2}$, Eq.~\eqref{eq:approxCT} perfectly matches Eq. \eqref{eq:DimerHC} up to higher temperatures. This would imply a universal scaling collapse in $C$ for power law $w(\Delta)$, as observed in Ref.~\cite{kimchi2018scaling}.

\section{Susceptibility from singlet distribution model}
\begin{figure}[b]
	\centering
	\includegraphics[width=\linewidth]{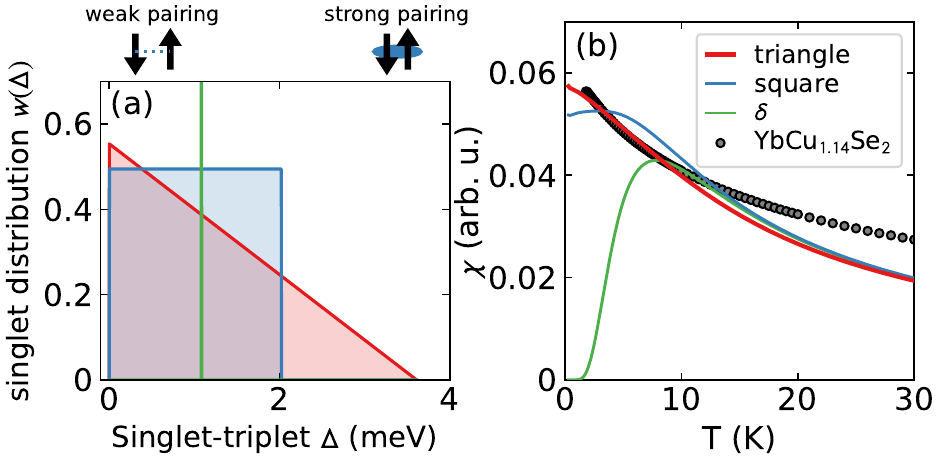}[h!]
	\caption{Calculated susceptibility for different distributions of singlets at $\mu_0 H = 0.1$~T, compared to experimental YbCu$_{1.14}$Se$_2$ susceptibility.}
	\label{fig:DimerSusceptibility}
\end{figure}

The singlet distribution model can also be used to compute susceptibility. The equation for dimer susceptibility with singlet-triplet gap $\Delta$ and at magnetic field $h$ is given by
\begin{eqnarray}
        \chi(\Delta,T,h) = \frac{4\mu^2}{k_B T} e^{-\Delta/k_B T} \frac{1}{Z}  \bigg( 2\cosh \Big(\frac{2\mu h}{k_B T}\Big) \nonumber \\ 
        - \frac{e^{-\Delta/k_B T}}{Z} 2 \sinh^2\Big(\frac{2\mu h}{k_B T}\Big) \bigg) 
\end{eqnarray}
where $\mu$ is the effective moment size of the spins and $Z$ is the partition function (this is a straightforward calculation from $\chi(T) = k_B T \frac{\partial^2 \ln Z}{ \partial h^2}$). We can sum over a distribution in the same way as Eq.~\ref{eq:SumSpecificHeat} summed over heat capacity. 
In Fig.~\ref{fig:DimerSusceptibility} we compute the susceptibility of distributions of dimers from the main text. Just as with $C/T$, the low temperature limit of susceptibility follows the same trend as the distribution: the triangular distribution gives linear $\chi$ as $T \rightarrow 0$, the square distribution gives a flat $\chi$ as $T \rightarrow 0$, and the delta function gives $\chi = 0$ as $T \rightarrow 0$. 
(As an aside, this highlights how a power law distribution of singlet splitting leads to the scaling collapse in Refs.~\cite{Kimchi_2018_YMGO,kimchi2018scaling}.)
Unlike heat capacity however, it is not as straightforward to compare to the experimental data because the crystal electric fields cause the effective moment to change as a function of temperature, and having the correct moment (slope in $\chi^{-1}$) at low temperatures causes the model to have too small susceptibility at high temperatures, which is seen in Fig.~\ref{fig:DimerSusceptibility}. 

Absent a reliable CEF model, we merely focus on the lowest temperature behavior ($T < 5$~K). The triangular distribution matches the experimental data much better than a flat distribution or a delta function distribution, though the overall size of the moment was scaled as a free parameter. This confirms that the triangular distribution of singlets is a reasonable approximation to the magnetic state in YbCu$_{1.14}$Se$_2$. 

\section{Sample-dependent specific heat}
\begin{figure}[b]
	\centering
	\includegraphics[width=0.85\linewidth]{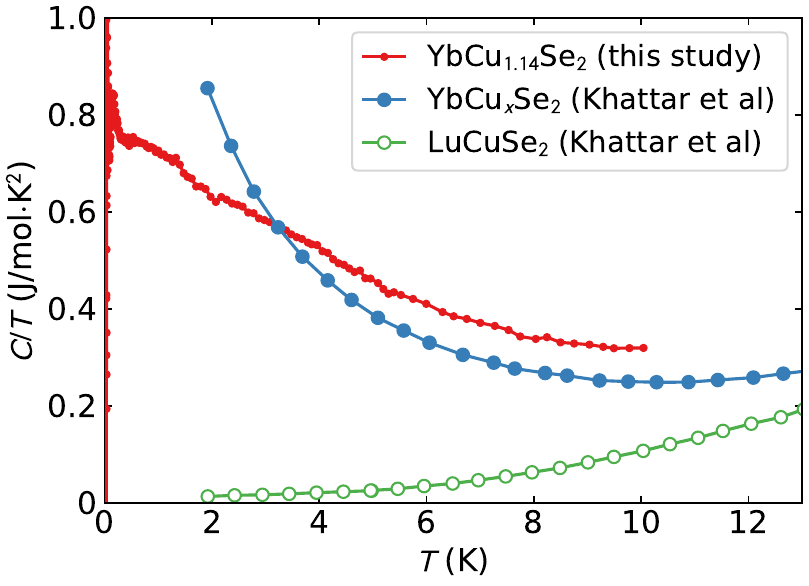}
	\caption{Specific heat for YbCu$_x$Se$_2$ from this study and from Ref.~\cite{khattar2025magnetic} alongside LuCuSe$_2$ data from Ref.~\cite{khattar2025magnetic}. The low temperature data do not match, indicating differences in the magnetic density of states from competing synthesis methods.}
	\label{fig:YbLu}
\end{figure}

In the final stages of manuscript preparation, Ref.~\cite{khattar2025magnetic} appeared which also reports the basic properties of YbCuSe$_2$. 
In Fig.~\ref{fig:YbLu} we plot the heat capacity from Ref.~\cite{khattar2025magnetic} compared with the data in this study. The discrepancy is quite dramatic, with Ref.~\cite{khattar2025magnetic} following much more of a power law. This indicates a noticeable difference in the magnetic density of states from different synthesis processes, and highlights that sample quality and stoichiometry have substantial effects on the collective magnetism.

\bibliography{triangular}

\end{document}